\documentclass[12pt,showpacs,preprintnumbers,amsmath,amssymb]{revtex4}
\usepackage{dcolumn}
\usepackage{epsfig}

\newcommand{\be}{\begin{equation}}
\newcommand{\ee}{\end{equation}}
\newcommand{\bea}{\begin{eqnarray}}
\newcommand{\eea}{\end{eqnarray}}
\begin{document}

\title{Regular non-Abelian vacua in ${\cal N}=4$, SO(4) gauged supergravity }
\author{Ali H. Chamseddine$^\dagger$ and  Mikhail S. Volkov$^{\dagger\dagger}$ }
\affiliation{$\dagger$ CAMS and Physics Department, American University of Beirut, LEBANON\\
$\dagger\dagger$ LMPT CNRS-UMR 6083,
Universit\'e de Tours, Parc de Grandmont,
37200 Tours, FRANCE
}

\begin{abstract}
 {We present a family of globally regular ${\cal N}=1$ vacua  in the D=4, ${\cal N}=4$
gauged supergravity of Gates and Zwiebach. These solutions
are labeled by the ratio $\xi$ of the two gauge couplings,
and for $\xi=0$ they reduce to
 the supergravity monopole previously used for constructing the gravity dual
of  ${\cal N}=1$
super Yang-Mills theory.
For $\xi>0$ the
solutions are asymptotically anti de Sitter, but  with
an excess of the solid angle, and they reduce exactly to anti de Sitter for $\xi=1$.
Solutions with $\xi<0$ are topologically $R^1\times S^3$, and for $\xi=-2$ they become
$R^1\times S^3$ geometrically.
All solutions with $\xi\neq 0$ can be promoted to D=11 to become vacua of M-theory.}
\end{abstract}
\maketitle

Regular supersymmetric backgrounds in gauged supergravities (SUGRA) play an important role
in the context of the AdS/CFT correspondence (see \cite{ADSCFT} for a review).
Upon uplifting to higher dimensions they become vacua of string/M theory
and can be used for the dual description of
strongly coupled gauge field theories. In this way,
for example, the monopole  solution \cite{CV}  of the ${\cal N}=4$
gauged SUGRA  has given rise to the holographic
interpretation of confining
${\cal N}=1$ super Yang-Mills (SYM) theory
\cite{MN}.
Constructing such solutions, however, is rather involved. This is why,
despite their importance, very few regular vacua of gauged
SUGRA's are known.

In this Letter we present a family of globally regular ${\cal
N}=1$ vacua that contains the monopole solution of Ref. \cite{CV}
as special case. We work in the context of the ${\cal N}=4$ gauged
SUGRA in four dimensions. This theory  exists in two inequivalent
versions: the SU(2)$\times$SU(2) model of Freedman and Schwarz
(FS) \cite{FS}, whose solutions were studied in \cite{CV},
 and the SO(4)
model of Gates and Zwiebach (GZ) \cite{GZ}. Both models contain in the
bosonic sector the graviton $g_{\mu\nu}$, dilaton $\phi$, axion
${\bf a}$, and two non-Abelian gauge fields $A^a_\mu$ and
$B^a_\mu$ with gauge couplings $e_A$ and $e_B$ and with gauge
group SU(2)$\times$SU(2). The important difference between the two
models is that in the FS model the dilaton potential
 has no stationary points, while in the GZ model one has
(when ${\bf a}=0$) \be \label{0} {\rm U}(\phi)=-\frac{e_A^2}{8}\,
(e^{-2\phi}+\xi^2e^{2\phi}+4\xi). \ee This potential does have
stationary points, and,
 depending on the sign of $\xi\equiv e_B/e_A$,
its extremal value -- the cosmological constant --
%, $-\frac{e_A^2}{4}(2\xi+|\xi|)$,
can be positive or negative.  If one sets ${\bf a}=B^a_\mu=\xi=0$,
then the FS and GZ models coincide and admit as a solution the
${\cal N}=1$ vacuum of Ref. \cite{CV} -- the Chamseddine-Volkov
(CV) monopole. If ${\bf a}=B^a_\mu=0$  but $\xi\neq 0$, then  the
two models are no longer the same, and we find that within the GZ
model the CV monopole admits generalizations for any $\xi\neq 0$.
These solutions are topologically different from the CV monopole,
although approach the latter pointwise as $\xi\to 0$. They can be
uplifted to D=11, which may suggest a holographic interpretation
for them.

We consider the ${\bf a}=B^a_\mu=0$ truncation of the GZ model
whose bosonic sector is described by the Lagrangian \bea
\label{lag} {\cal L}  =\frac{1}{4}R
-\frac12\partial_\mu\phi\partial^\mu\phi
-\frac{1}{4}\,e^{2\phi}F^{a}_{\mu\nu}F^{a\mu\nu} -{\rm U}(\phi).
\eea Here $F^a_{\mu\nu}=\partial_\mu A^a_\nu-\partial_\nu A^a_\mu
+\epsilon_{abc}A^b_\mu A^c_\nu$ with $a=1,2,3$, the scale is
chosen such that $e_A=1$, and U$(\phi)$ is given by (\ref{0}).
Consistency of setting the axion to zero requires that $\ast
F^{a}_{\mu\nu}{F}^{a\mu\nu}=0$. The theory contains also fermions:
the gaugino $\chi$ and gravitino $\psi_\mu$, whose SUSY variations
for a purely bosonic background are \bea \label{SUSY}
\delta{\chi}&=& \frac{1}{\sqrt{2}}\,\gamma^\mu
\partial_\mu\phi\,\epsilon+ \frac{1}{2}\,e^{\phi}{\cal F}\epsilon
+\frac{1}{4}\,(e^{-\phi}-\xi e^{\phi})\epsilon, \\
\delta{\psi}_\mu&=&
{\cal D}_\mu\epsilon +
\frac{1}{2\sqrt{2}}\,e^{\phi}{\cal F}\gamma_\mu \epsilon
+\frac{1}{4\sqrt{2}}\,(e^{-\phi}+\xi e^{\phi})\gamma_\mu\epsilon.  \nonumber
\eea
Here ${\cal F}=\frac12\alpha^aF^a_{\alpha\beta}\gamma^\alpha\gamma^\beta$ and
$
{\cal D}_{\mu}=
{\partial}_{\mu}
+\frac14\,\omega_{\alpha\beta,\mu}\gamma^\alpha\gamma^\beta
-\frac{1}{2}A^a_{\mu}\alpha^a$;
the late ($\mu,\nu$) and early $(\alpha,\beta)$ Greek letters correspond
to the spacetime and tangent space indices, respectively.
The gamma matrices are subject to
$\frac12(\gamma_{{\alpha}}\gamma_{{\beta}}
+\gamma_{{\beta}}\gamma_{{\alpha}})=
\eta_{\alpha\beta}\equiv
{\rm diag}(-1,1,1,1)$.
Introducing Pauli matrices of four different types,
$\sigma^a$, $\underline{\sigma}^b$, $\tau^c$, $\underline{\tau}^d$,
which act in four different spaces, respectively
(such that, for example, $[\sigma^a,\underline{\sigma}^b]=0$),
one can choose $\gamma^\alpha\equiv
(\gamma^0,\gamma^a)=(i\underline{\sigma}^1,\underline{\sigma}^2\sigma^a)$.
The gauge group SU(2)$\times$SU(2) is generated by the antihermitean matrices
$\alpha^a$ and $\beta^b$, $[\alpha^a,\beta^b]=0$,
$\alpha^a\alpha^b=-\epsilon_{abc}\alpha^c-\delta_{ab}$,
and similarly for $\beta^a$. One can choose
$\alpha^a=i\tau^a$, $\beta^a=i\underline{\tau}^a$.
The generators $\beta^a$ correspond to the field $B^a_\mu$
that is truncated to zero.

We wish to study fields that preserve some of the supersymmetries,
in which case $\delta\chi=\delta\psi_\mu=0$ for certain $\epsilon\neq 0$.
We restrict to
the static and spherically
symmetric sector parameterized by coordinates $(t,\rho,\vartheta,\varphi)$
with
\bea                                          \label{sol}
ds^2_{(4)}&=&-e^{2V(\rho)}dt^2+e^{2\lambda(\rho)}d\rho^2+r^2(\rho)d\Omega^2, \\
\tau^a A^a_\mu dx^\mu&=&\frac{i}{2}(1-w(\rho))[T,dT],~~~~~\phi=\phi(\rho).   \label{sol1}
\eea
Here, with $n^a=(\sin\vartheta\cos\varphi,\sin\vartheta\sin\varphi,\cos\vartheta)$, one has
 $d\Omega^2=dn^a dn^a$ and $T=\tau^a n^a$. Imposing the isotropic
gauge condition, $r=\rho e^{\lambda}$, the spatial part of the metric becomes
conformally flat, $ds_{(3)}^2=e^{2\lambda}dx^a dx^a$, with $x^a=\rho n^a$.
Choosing the
tetrad $\theta^\alpha=(e^{V}dt,e^{\lambda}dx^a)$,
the spin connection is obtained from
$d\theta_\alpha+\omega_{\alpha\beta}\wedge\theta^\beta=0$.
Setting in (\ref{SUSY})  $\delta\chi=\delta\psi_\mu=0$
gives then the equations for the
SUSY Killing spinors $\epsilon$:
\bea                   \label{d0}
0&=&{2\sqrt{2}}\,{\rm e}^{-\lambda}\phi'\,
\underline{\sigma}^2\,
(\vec{n}\vec{\sigma})\epsilon+
2\,{\rm e}^{\phi}\,
{\cal F} \epsilon
+
({\rm e}^{-\phi}-\xi e^\phi)
\epsilon \, ,              \nonumber          \\
0&=&2i\sqrt{2}e^{-V}\underline{\sigma}^1\partial_t\epsilon+
{\sqrt{2}}\,e^{-\lambda}(V-\phi)'\underline{\sigma}^2(\vec{n}\vec{\sigma}) \epsilon
+\xi e^\phi\epsilon,                   \nonumber      \\
0&=&
\vec{\nabla}\epsilon+
\frac{i}{2}\,\lambda'\,
(\vec{n}\times\vec{\sigma})\epsilon
+\frac{i}{2}\frac{w-1}{\rho}\,
(\vec{n}\times\vec{\tau})\epsilon
\nonumber      \\
&+&
\frac{e^{\lambda}}{4\sqrt{2}}(
2{\rm e}^{\phi}{\cal F}+
{\rm e}^{-\phi}+\xi e^\phi)\underline{\sigma}^2
\vec{\sigma}\epsilon.
\eea
Here
${\cal F}=-r^{-2}\, [
 \rho w'(\vec{\sigma}\vec{\tau})+
 (w^2-1-\rho w')
 (\vec{n}\vec{\sigma})(\vec{n}\vec{\tau})]$ and $^{\prime}\equiv \frac{d}{d\rho}$;
also  the usual operations for Euclidean 3-vectors
are assumed, for example
$\vec{n}\vec{\sigma}\equiv n^a\sigma^a$ and
$\vec{\nabla}\equiv \partial/\partial x^a$.
Eqs. (\ref{d0}) comprise
an overdetermined system of 80 equations for 16 components of $\epsilon$,
whose consistency conditions we shall now study.

Let $\psi_A$, $\chi_A$, $\underline{\psi}_A$,
$\underline{\chi}_A$ be eigenspinors of $\sigma^3$, $\tau^3$,
$\underline{\sigma}^1$, $\underline{\tau}^2$, respectively,
with the eigenvalues $(-1)^A$, $A=1,2$.
We make the ansatz
$\epsilon=\epsilon_{AB}$ with $A,B=1,2$ and
\be                                  \label{spinor}
\epsilon_{AB}=
{\cal U}\exp(i\Psi t)
[\Phi_{-}(\rho)+
\Phi_{+}(\rho)\,\underline{\sigma}^2
(\vec{n}\vec{\sigma})]\,
\epsilon_0\,\underline{\psi}_A\underline{\chi}_B.
\ee
Here
$\Psi=\frac{1}{2\sqrt{2}}(-1)^{A+B}{\xi Q\, }\,\underline{\tau}^2$
where $Q$ is a real constant,
${\cal U}=\exp(-\frac{i}{2}\tau_3\varphi)\exp(-\frac{i}{2}\tau_2\vartheta)$,
and
$\epsilon_0=\psi_1\,\chi_2-\psi_{2}\chi_1$.  In fact, $ \epsilon_{AB}$
is the most general spinor whose total angular momentum,
including the orbital part plus spin plus isospin, is zero.
 Inserting (\ref{spinor})  into (\ref{d0}), the variables decouple, and the
system reduces to six linear algebraic and two ordinary
differential equations for $\Phi_{\pm}$: 
\bea \label{dif}
{A}^{\pm}_{(m)}\Phi_\pm=B^{\mp}_{(m)}\Phi_\mp,~~~~
\Phi_{\pm}'-{\cal A}\Phi_\pm ={\cal B}\Phi_\mp~. 
\eea 
Here the
coefficients ${A}^{\pm}_{(m)}$, ${B}^{\pm}_{(m)}$ ($m=1,2,3$),
${\cal A}$ and ${\cal B}$ are functions of the background
amplitudes $V,\lambda,r,w,\phi$ and their derivatives. The
algebraic equations can have a nontrivial solution if only their
coefficients fulfill the conditions $A^{+}_{(m)}
A^{-}_{(n)}=B^{+}_{(m)} B^{-}_{(n)}$, of which only 5 are
independent. Introducing $N=\rho\lambda'+1$, these 5 conditions
are equivalent to the first 5 of the  following 6 relations: \bea
&&V'-\phi'=\xi\,\frac{P}{\sqrt{2}N}e^{\phi+\lambda},~~~~~
Q=e^{V+\phi}\frac{w}{N}, ~~~~~\label{1}  \\
&&\phi'=\sqrt{2}\,\frac{BP}{N}\, e^{\lambda},~~~~~~
w'=-\frac{rwB}{N} e^{-\phi+\lambda}, \label{2} \\
&&N=\sqrt{w^2+P^2},~~~~~~
r'=Ne^\lambda.   \label{3}
\eea
Here
$P=e^\phi\frac{1-w^2}{\sqrt{2}r}+\frac{r}{2\sqrt{2}}(e^{-\phi}+\xi e^\phi)$
and $B=-\frac{P}{\sqrt{2}r}+\frac12e^{-\phi}$.
These relations impose nonlinear differential constraints on the
background functions $\phi$, $w$, $V$, $\lambda$ and the parameter $Q$.
Remarkably, although we have in (\ref{1}) two equations for the
same function $V(\rho)$, the first of these equations is in fact a differential
consequence of the second one, and so the system is not overdetermined.
The last equation in (\ref{3}), added for the later convenience,
is the identity (in the isotropic gauge used) implied by the definition
of $N$.
If Eqs.(\ref{1})--(\ref{3})
are fulfilled, the algebraic equations in (\ref{dif})
are consistent with each other and express $\Phi_{-}$ in terms of $\Phi_{+}$.
Inserting this to the first
differential constraint in (\ref{dif}) gives a linear differential equation
for $\Phi_{+}(\rho)$, whose solution can be expressed in quadratures.
The second differential constraint in (\ref{dif}) then turns out to be fulfilled
{ identically}, by virtue of Eqs. (\ref{1})--(\ref{3}).

The Bogomolnyi equations  (\ref{1})--(\ref{3}) therefore provide
the full set of consistency conditions that guarantee the
existence of SUSY Killing spinors.  One can now pass in these
equations to an arbitrary gauge by treating $\lambda(\rho)$ as a
free function subject to a gauge condition, while considering the
second relation in (\ref{3}) as the dynamical equation for the
Schwarzschild radial function $r(\rho)$.
 Finding then $\Phi_{\pm}$ gives the spinor $\epsilon_{AB}$ for each choice of $A,B$,
which finally corresponds to four independent SUSY Killing spinors, that is to
${\cal N}=1$.

Introducing $y^1=w$, $y^2=\phi$, $y^3=V$, $y^4=V+\ln r$, the
Bogomolnyi equations can also be written as \be \label{B}
Y^n\equiv \frac{dy^n}{d\rho}-G^{nm}\frac{\partial {\cal
W}}{\partial y^m}=0,~~~~~ m,n=1,2,3,4, \ee where the target space
metric is defined by $G_{mn}=r^2\,e^{V-\lambda}{\rm
diag}(2e^{2\phi}/r^2,1,1,-1)$ and the superpotential is ${\cal
W}=-r e^V N$ with $N$ given by (\ref{3}). We note also that
inserting the ansatz (\ref{sol}),(\ref{sol1}) to the Lagrangian
(\ref{lag}) and integrating over the angles gives $\int {\cal
L}\sqrt{-g}\,d\vartheta d\varphi=4\pi G_{mn}Y^mY^n+$total
derivative. It then follows that solutions of Eqs. (\ref{B}) are
stationary points of the action.

To integrate the Bogomolnyi equations (\ref{1})--(\ref{3}), the
problem actually reduces to studying the closed subsystem
(\ref{2})--(\ref{3}) for $\phi,w,r$, since $V$ can be obtained
afterwards from (\ref{1}). It seems that these equations can be
resolved analytically only for some special values of $\xi$, and
we shall therefore resort to numerical analysis  to study the
generic case. First of all, we notice the following symmetry of
the equations: if $\xi$, $\phi(\rho)$, $w(\rho)$, $r(\rho)$,
$\lambda(\rho)$ is a solution  for some value of $\xi$, then, for
any $\epsilon$, 
\be \label{sym}
e^{-2\epsilon}\xi,~~\phi(\rho)+\epsilon,~~w(\rho),~~e^\epsilon
r(\rho), ~~\lambda(\rho)+\epsilon 
\ee 
is also a solution. To fix
this symmetry, we impose the condition $\phi_0=0$, with $\phi_0$
being the value of $\phi$ at $r=0$. Since  the sign of $\xi$ is
invariant under (\ref{sym}), there are three separate cases to
study: $\xi>0$, $\xi<0$, and $\xi=0$.

\underline{$\xi=0$.} Eqs. (\ref{2})--(\ref{3}) reduce in this case
to the system previously studied  in the context of the
half-gauged FS model \cite{CV}. Its solution is the CV monopole:
\bea                                                   \label{CV}
w=\frac{\rho}{\sinh\rho}, ~~~~~
\frac{re^{-\phi}}{\sqrt{2}}=\frac{\rho
e^{-2\phi}}{w}=\sqrt{2\rho\coth\rho-w^2-1},~~ 
\eea and
$e^\lambda=\sqrt{2}e^\phi$. In this case one has $\phi\sim\rho$ as
$\rho\to\infty$.

%%%%%%%%%%%%%%%%%%%%%%%%%%%%%%%%%%%%
\begin{figure}
\epsfxsize=13cm
\epsfysize=8cm
\centerline{\epsffile{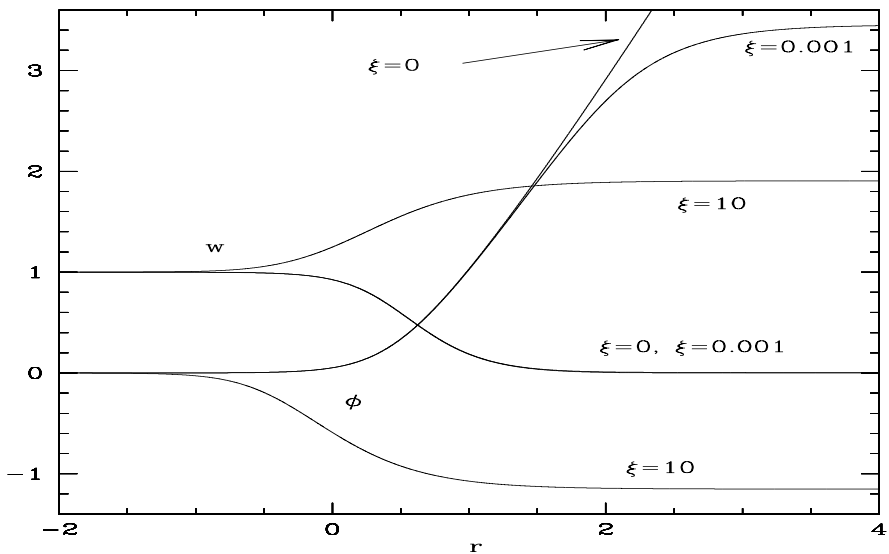}}
\caption{Globally regular solutions with $\xi>0$.  For $\xi=0.001$ and $\xi=0$
the amplitudes $w$ are almost identical.}
\label{fig1}
\end{figure}
%%%%%%%%%%%%%%%%%%%%%%%%%%%%%%%%%%%%
\underline{$\xi>0$.}
Choosing the Schwarzschild gauge, $\rho=r$,
the essential equations are given by (\ref{2}),(\ref{3})
with $e^\lambda=1/N$.
They determine $\phi(r)$, $w(r)$,
 while (\ref{1}) gives $e^{2V}=Q^2{N^2}e^{-2\phi}/w^2$.
We are interested in everywhere regular solutions, in which case
$\phi=O(r^2)$,  $w=1+O(r^2)$, $N=1+O(r^2)$ 
 for $r\to 0$. For $r\to\infty$ one has
\bea                        \label{infinity}
\phi&=&-\frac{\ln\xi}{2}+\frac{b}{r} +O(r^{-2}),~~~
w=w_\ast-\frac{w_\ast b}{r}  +O(r^{-2}), \nonumber  \\
 N^2&=&1+\frac12\,\xi b^2-\frac{2b(w_\ast^2-1)}{r}+\frac{\xi r^2}{2}+O(r^{-2}), 
\eea
where $b$ and $w_\ast$ are integration constants.
The numerical integration of the equations  reveals
for every value of $\xi>0$ a global  solution $\phi(r)$, $w(r)$ with such boundary
conditions in the interval
$r\in[0,\infty)$; see Fig.I. For all these solutions the dilaton varies in the
finite range and runs into the stationary point of U$(\phi)$ for $\rho\to\infty$.
As $\xi\to 0$, the asymptotic value of $\phi$ tends to infinity and
the solutions approach pointwise
the CV monopole  (\ref{CV}).

For $\xi=1$ the solution can be obtained
analytically: $\phi(r)=0$, $w(r)=1$.
Choosing $Q=1$ (the value of $Q$ can be adjusted
by rescaling the time), 
the metric assumes the
 standard anti de Sitter (AdS) form,
$ds^2=-N^2dt^2
+dr^2/N^2
+r^2d\Omega^2$
with $N^2=1+\frac{r^2}{2}$,
while the gauge field vanishes.
This solution actually has ${\cal N}=4$ supersymmetry,
since in this case there are additional SUSY Killing spinors
not contained in the ansatz (\ref{spinor}).

Solutions with $\xi\neq 1$ describe globally regular ${\cal N}=1$
deformations of the AdS.
Their asymptotic form is determined by (\ref{infinity}). Choosing the new radial
coordinate $\tilde{r}=r/\sqrt{1+\delta}$ with $\delta=\frac12\,\xi b^2$
and setting $Q=w_\ast/\sqrt{\xi(1+\delta)}$,
the metric asymptotically approaches
\be
ds^2=-N^2  dt^2
+\frac{d\tilde{r}^2}{N^2}
+(1+\delta)\,\tilde{r}^2d\Omega^2,
\ee
where $N^2=1-\frac{2M}{\tilde{r}}+\frac{\xi\tilde{r}^2}{2}$ and
$M=b(w_\ast^2-1)/(1+\delta)^{3/2}$.
This is the Schwarzschild-AdS metric with an {\it excess} of the solid angle --
the area of the 2-sphere of constant $\tilde{r}$ being $4\pi(1+\delta)\tilde{r}^2$
in this geometry. The excess parameter $\delta$ and the `mass' $M$
vanish only for $\xi=1$, and they tend to infinity as $\xi\to 0$.

\underline{$\xi<0$.} Solutions in this case are of the `bag of gold' type, since
they have compact spatial sections with the $S^3$ topology.
The range of the Schwarzschild function $r(\rho)$ is finite:
it starts from zero at $\rho=0$ (`north pole'), increases up to a maximal
value at some $\rho_e>0$ (`equator'), and then decreases down to zero at some
$\rho_\ast>\rho_e$ (`south pole'). Since
$N\sim r'=0$ at the equator, Eqs. (\ref{1})--(\ref{3})
become singular at this point. To desingularize them, we set
$P=wS$ thus obtaining $N=w\sqrt{1+S^2}=we^{V+\phi}$
(having chosen $Q=1$ in (\ref{1})). Using this in (\ref{1})--(\ref{3}) reduces
the system to
\bea                           \label{reg}
r'&=&we^{V+\phi+\lambda},~~~~~\phi'=\sqrt{2}BSe^{\lambda-V-\phi}, \\
w'&=&-rBe^{\lambda-V-2\phi},~~~~~V'-\phi'=\frac{\xi}{\sqrt{2}}Se^{\lambda-V},
\nonumber
\eea
with  $S=\sqrt{e^{2V+2\phi}-1}$
and $B=-\frac{wS}{\sqrt{2}r}+\frac12 e^{-\phi}$. In addition, the relation
$P=wS$ with $P$ defined after Eqs. (\ref{3}) gives the first integral
for these equations.
Eqs. (\ref{reg}) are completely regular at the equator,
whose position coincides with zero of $w$.
Imposing the gauge condition
$\lambda=0$ and demanding the solution to be regular at the `north pole' gives
$r=\rho+O(\rho^3)$, $w=1+O(\rho^2)$, $\phi=O(\rho^2)$, $V=O(\rho^2)$
for small $\rho$. At the `south pole' we find the formal
power series solution to be generically
\bea                                                           \label{pole}
r=3w_\ast\beta_\ast\,x+O(x^3),~
w=w_\ast + \frac{w_\ast \kappa}{8} \,x^4+O(x^6),~  \\
e^{-\phi}=\frac{|\kappa| x}{3\sqrt{2}}
+O(x^3), ~
e^{V-\phi}=\nu_\ast +\frac{\xi\beta_\ast\kappa}{4}\,x^2+O(x^4),
\nonumber
\eea
with $x=(\rho-\rho_\ast)^{1/3}$. Here $\rho_\ast$, $w_\ast<0$,
$\beta_\ast>0$ and $\nu_\ast$ are integration constants, and
$\kappa=(1-w_\ast^2)/(w_\ast^2 \beta_\ast^2)$.

Solutions of Eqs. (\ref{reg}) in the interval $\rho\in[0,\rho_\ast]$
comprise a one-parameter family labeled by $\xi$.
These solutions are regular for $\rho<\rho_\ast$,
while at $\rho=\rho_\ast$
the dilaton diverges and the curvature is singular too. For any $\xi$, the
profile of these solutions is qualitatively similar to the one shown in Fig. 2.
As $\xi\to 0$, one has $\rho_\ast\to\infty$,  $r(\rho_e)\to\infty$,
and the solutions approach pointwise
the CV monopole.
%%%%%%%%%%%%%%%%%%%%%%%%%%%%%%%%%%%%
\begin{figure}
\epsfxsize=13cm
\epsfysize=8cm
\centerline{\epsffile{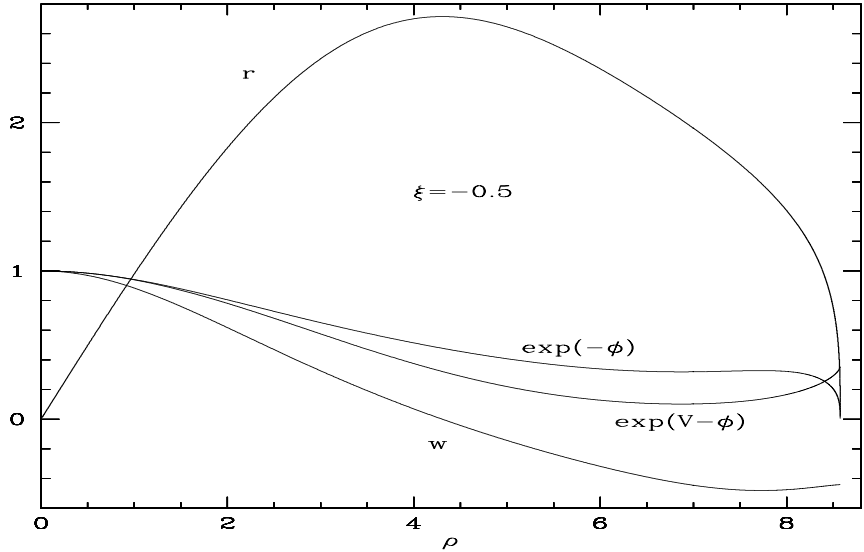}}
\caption{The compact solutions with $\xi<0$. They are generically
singular at the `south pole' where $r$ vanishes and $\phi$  diverges.
}
\label{fig2}
\end{figure}
%%%%%%%%%%%%%%%%%%%%%%%%%%%%%%%%%%%%

For one special value,  $\xi=-2$,
 one has $w_\ast=-1$ and the expansions (\ref{pole}) are no longer valid.
However, the solution can then be obtained analytically:
$\phi=V=P=S=0$, $w=\cos\rho$,
$r=\sqrt{2}\sin\rho$, $e^{\lambda}=\sqrt{2}$.
This solution is {\it globally} regular, also at the south pole,
the spatial geometry being
that of the round $S^3$.
One can write down the metric and gauge field as
\be                                                 \label{-2}
ds^2=-dt^2+2\theta^a\theta^a,~~~~~~A^a=\theta^a,
\ee
where $\theta^a$ are invariant forms
on $S^3$, $d\theta^a+\epsilon_{abc}\theta^b\wedge\theta^c=0$.
However, there is no SUSY enhancement in this case, and so
${\cal N}=1$.

We have thus obtained the generalizations of the CV monopole (\ref{CV})
that comprise a two-parameter family labeled by $\xi$
and $\phi_0$. Although we have described explicitly only
solutions with $\phi_0=0$, those with $\phi_0\neq 0$ can be obtained by using the
symmetry (\ref{sym}). The solutions generically have ${\cal N}=1$, while for
$\xi=1$ the sypersymmetry is enhanced up to  ${\cal N}=4$.
We know that the $\xi=0$ solution can be uplifted to D=10
to become a vacuum of string theory \cite{CV}.
It turns out that solutions with $\xi\neq 0$ can be
uplifted to $D=11$ to become vacua of M-theory.

The derivation of the GZ model via dimensional reduction of D=11 SUGRA
was considered in Ref. \cite{Pope}. Using formulas given in there combined 
with the symmetry  (\ref{sym}), 
every D=4 vacuum  $(ds_4^2,A^a,\phi)$ considered above
maps to the M-theory solution $(ds_{(11)}^2,F_{[4]})$.
The metric in D=11 is given by
$ds^2_{(11)}=|\xi|\Delta^\frac{2}{3}ds_4^2+8\Delta^{-\frac13}ds_{(7)}^2$
with
\be                                            \label{s7}
ds^2_{(7)}=
\Delta\, d\eta^2
+\frac{{\bf c}^2}{X} \sum_a(\theta^a_{(1)}-\frac12 A^a)^2
+{{\bf s}^2}X\sum_a(\theta_{(2)}^a)^2.
\ee
Here
$\Delta={\bf s}^2/X+{\bf c}^2 X$ with $1/X=\sqrt{|\xi|}e^{\phi}$, and
$\theta^a_{(\iota)}$  $(\iota=1,2)$ are invariant forms on two different
3-spheres,
$d\theta^a_\iota+\epsilon_{abc}\theta^b_\iota\wedge \theta^c_\iota=0$.
The case $\xi>0$ corresponds to the reduction on $S^7$,
one has then ${\bf c}=\cos\eta$, ${\bf s}=\sin\eta$ with $\eta\in[0,\pi/2]$,
while for $\xi<0$ one reduces on the ${\cal H}^{(2,2)}$ hyperbolic space \cite{Pope},
in which case ${\bf c}=\cosh\eta$, ${\bf s}=\sinh\eta$ and $\eta\in[0,\infty)$.
The 4-form in D=11 reads
\be                               \label{F4}
F_{[4]}=\sqrt{2}{\bf sc} \ast d\phi\wedge
d\eta -\left(
\frac{\xi}{|\xi|}\,{{\bf c}^2}X^2+\frac{{\bf s}^2}{X^2}+2
\right)\frac{\epsilon_{(4)}}{\sqrt{2}}, \ee
where $\epsilon_{(4)}$  is the 4-volume form and $\ast$ is the Hodge dual in the
4-space.

These formulas suggest a holographic interpretation for our solutions.
For $\xi=0$,  according to \cite{MN}, the dual theory is
 $D=4$, ${\cal N}=1$ SYM.  For $\xi=1$
we have $\phi=A^a=0$,  and the D=11 geometry
is $AdS_4\times S^7$. This is the near-horizon limit of the M2 brane,
and the dual theory is therefore $D=3$, ${\cal N}=4$ SYM. This suggests that other solutions
with positive $\xi\neq 1$ may describe some ${\cal N}=1$ deformations of this theory.
It would also be interesting to work out an interpretation for the compact solutions
with $\xi<0$,
especially for the one given by Eq. (\ref{-2}).

M.S.V. thanks Michaela Petrini for discussions. Research of A.H.C.
is supported in part by National Science Foundation Grant No.
Phys-0313416.

\end{document}